\documentclass[useAMS,usenatbib]{mn2e}
\usepackage{graphicx}
\usepackage{amsfonts}
\usepackage{amsmath}
\usepackage{color}
\usepackage{soul}

\newif\ifAMStwofonts
\AMStwofontstrue

\def\gs{\mathrel{\raise0.35ex\hbox{$\scriptstyle >$}\kern-0.6em 
\lower0.40ex\hbox{$\scriptstyle \sim$}}}
\def\ls{\mathrel{\raise0.35ex\hbox{$\scriptstyle <$}\kern-0.6em 
\lower0.40ex\hbox{$\scriptstyle \sim$}}}

\newcommand{\msun}{M_\odot}

\newcommand{\be}{\begin{equation}}
\newcommand{\ee}{\end{equation}}
\newcommand{\bea}{\begin{eqnarray}}
\newcommand{\eea}{\end{eqnarray}}

\def\gsim{\lower.5ex\hbox{\gtsima}}
\def\lsim{\lower.5ex\hbox{\ltsima}}
\def\gtsima{$\; \buildrel > \over \sim \;$}
\def\ltsima{$\; \buildrel < \over \sim \;$}

\def\gsim{\lower.5ex\hbox{\gtsima}}
\def\lsim{\lower.5ex\hbox{\ltsima}}
\def\simgt{\lower.5ex\hbox{\gtsima}}
\def\simlt{\lower.5ex\hbox{\ltsima}}

\ifCUPmtlplainloaded \else
  \ifAMStwofonts \else 
    \def\upi{\pi}
                 \def\umu{\mu}
    \def\upartial{\partial}
  \fi
\fi

\title{Consequences of dark matter self-annihilation for galaxy 
formation}

\author[Natarajan, Croton \& Bertone]{Priyamvada 
Natarajan$^{1,2}$, Darren Croton$^{3}$ \& Gianfranco Bertone$^{4}$\\
$^1$Department of Astronomy, Yale University, P.O. Box 208101, 
       New Haven, CT 06520-8101, USA \\
$^2$Department of Physics, Yale University, P. O. Box 208120,
New Haven, CT 06520-8120, USA\\
$^3$Department of Astronomy, University of California, Berkeley, 
CA 94720, USA\\
$^4$Institut d'Astrophysique de Paris, UMR 7095-CNRS,
Universit\'{e} Pierre et Marie Curie, 98bis Bd Arago, 75014 Paris, France}

\pagerange{\pageref{firstpage}--\pageref{lastpage}}
\pubyear{2006}

\begin{document}

\maketitle

\label{firstpage}

\begin{abstract}
Galaxy formation requires a process that continually heats gas and
quenches star formation in order to reproduce the observed shape of
the luminosity function of bright galaxies. To accomplish this,
current models invoke heating from supernovae, and energy injection
from active galactic nuclei. However, observations of radio-loud
active galactic nuclei suggest that their feedback is likely to not be
as efficient as required, signaling the need for additional heating
processes. We propose the self-annihilation of weakly interacting
massive particles that constitute dark matter as a steady source of
heating. In this paper, we explore the circumstances under which this
process may provide the required energy input. To do so, dark matter
annihilations are incorporated into a galaxy formation model within
the Millennium cosmological simulation. Energy input from
self-annihilation can compensate for {\bf all} the required gas
cooling and reproduce the observed galaxy luminosity function only for
what appear to be extreme values of the relevant key parameters. The
key parameters are: the slope of the inner density profile of dark
matter haloes and the outer spike radius. The inner density profile
needs to be steepened to slopes of $-1.5$ or more and the outer spike
radius needs to extend to a few tens of parsecs on galaxy scales and a
kpc or so on cluster scales. If neutralinos or any thermal relic WIMP
with s-wave annihilation constitute dark matter, their
self-annihilation is inevitable and could provide enough power to
modulate galaxy formation. Energy from self-annihilating neutralinos
could be yet another piece of the feedback puzzle along with
supernovae and active galactic nuclei.
\end{abstract}


\section{Introduction}
\label{section1}

A major challenge to our current understanding of structure and galaxy
formation in the Universe is the discrepancy between the theoretically
predicted mass function of dark matter haloes and the observed shape
of the luminosity function of galaxies (Kauffmann \& White 1993; Cole
et al. 1994; Somerville, Primack \& Faber 2001; Benson et
al. 2003). Simply put, the number density of dark matter haloes falls
off as a power-law at high masses, whereas the luminosity function of
galaxies occupying such haloes terminates exponentially above a
characteristic luminosity (e.g. Cole et al. 2001; Huang et
al. 2003). This implies that the supply of gas to a galaxy and the
conversion of this gas into stars becomes preferentially inefficient
in more massive systems (White \& Rees 1978). A key issue for galaxy
formation theory is thus to illuminate the physical processes that
heat and cool gas in massive galaxies as this cycle regulates the
formation of new stars. In addition, one of the challenges for galaxy
formation is to unravel the inter-play of baryons with the
ubiquitous dark matter in galaxies.

To this end, feedback processes operating in galaxies at both the low
and high-mass ends of the halo mass function are required to explain
the faint and bright-end slopes of the observed galaxy luminosity
function. Energy input from supernovae is thought to play a
significant role in the regulation of star formation in low mass
galaxies (Dekel \& Silk 1986; Mac Low \& Ferrara 1999), however the
supernova energy injected in high mass galaxies is too small to
suppress gas cooling effectively. For such objects, to reconcile 
theory with observations a more energetic process that continually
heats the gas and operates independently of star formation appears to
be required (Croton et al. 2006; Bower et al. 2006).

Recent observations in nearby galaxies reveal a correlation between
the masses of supermassive black holes and the velocity dispersion of
the stellar component.  This suggests that black holes might play a
role in regulating star formation (Magorrian et al. 1998; Tremaine et
al. 2002; Ferrarese \& Merritt 2002). Energy input from nuclear
outflows driven by accreting black holes are currently favoured as the
principal source of feedback driving the truncation of star formation
in massive galaxies (Di Matteo et al. 2005; Croton et al. 2006; Bower
et al. 2006). Unlike supernovae feedback, the observed energy output
from Active Galactic Nuclei (AGN) can far exceed that liberated from
cooling gas in the hot halo. AGN heating in two distinct forms has
been proposed.  Episodic feedback from the AGN via outflows generated
during the merger process have been proposed by Di Matteo et
al. (2005); Sijacki \& Springel (2006), the so-called
`quasar-mode'. However, it is known that AGN spend most of their
lifetimes in a low accretion rate state, and therefore Croton et
al. (2006) and Bower et al. (2006) argue for a more steady, so called
`radio-mode' feedback that is long-lived.  The details of both these
processes are complex and the micro-physics is currently not well
understood.

Here, we focus on an alternative heating mechanism, the energy
steadily generated by the self-annihilation of dark matter particles
in the inner regions of haloes.  Although the standard cosmological
paradigm is predicated on the existence of non-baryonic dark matter
(DM) particles, the precise nature of these particles and their
interactions remain a puzzle. The neutralino is the current leading
dark matter candidate. In this paper we explore if the energy supplied
from the self-annihilation of dark matter in the centres of galaxies
and cluster haloes could possibly play a significant role in the
baryonic cooling/heating cycle (Ascasibar 2007; Totani 2004; 2005;
Colafrancesco et al. 2006). Detailed astrophysical implications of
neutralino dark matter annihilations in galaxy clusters, with a
specific application to the Coma cluster have been calculated by
Colafrancesco, Profumo \& Ullio (2006).  They performed a thorough
analysis of the transport and diffusion properties of neutralino
annihilation products, and investigated the resulting multi-frequency
signals, from radio to gamma-ray frequencies. They also study other
relevant astrophysical effects of neutralino annihilations, like the
DM-induced Sunyaev-Zel'dovich effect and the intracluster gas heating.

Annihilation luminosity is expected to be produced as a result of an
enhancement in the density of the dark matter distribution in the
inner-most regions of galaxies, due to the response of the dark matter
to the adiabatic growth of a central black hole (Gondolo \& Silk 1999)
and/or to the adiabatic compression suffered by the dark matter due to
the collapse of baryons into the dark matter potential well
(Blumenthal et al. 1986; Ryden \& Gunn 1987; Gnedin et al. 2004). By
including this steady heating source of annihilating neutralinos to
suppress cooling flow gas in a sophisticated model of galaxy
formation, we explore the effect of various key parameters in
reproducing the bright-end of the observed galaxy luminosity
function. The question we address is whether self-annihilation can
provide an alternative mechanism to AGN driven feedback within the
context of the current paradigm.

The outline of this paper is as follows. In Section~\ref{section2} we
discuss the physics of neutralino annihilation; a scenario for the
coupling of this energy source to baryons in galactic nuclei is
explored in Section~\ref{section3}. With this phenomenology developed,
we incorporate this feedback scheme into a galaxy formation model. The
details of the N-body simulation and galaxy formation model used in
this implementation are discussed in Section~\ref{section4}. Our
results are presented in detail in Section~\ref{section5} and
summarized in Section~6. We conclude with a discussion of the
implications of our results and comparison of annihilation heating
with other modes of feedback. Unless otherwise stated we assume Hubble
constant parametrised as $H_0=h\,100\,{\rm km}\,{\rm s}^{-1}\,{\rm
Mpc}^{-1}$ with $h=0.73$.

\section{Dark Matter candidates}
\label{section2}

Theoretical predictions of structure formation and evolution in a Cold
Dark Matter (CDM) cosmogony appear to be in very good agreement with
current observations on most scales bar perhaps the smallest (Spergel
et al.  2006; Tegmark et al. 2004; Seljak et al. 2005; Cole et
al. 2005).  In this well established paradigm, the bulk of the matter
in the Universe is comprised of cold, collisionless particles that
seed the formation of structures from the gravitational amplification
of their early fluctuations (Blumenthal et al. 1984; Davis,
Efstathiou, Frenk \& White 1985).

Weakly Interacting Massive Particles (WIMPs) are the currently
favoured dark matter candidates as their cross-sections are small
enough that they act as essentially collisionless particles (Kolb \&
Turner 1990). The super-symmetric {\it neutralino}, arising in minimal
super-symmetric extensions of the Standard Model of particle physics,
is probably the most widely studied WIMP candidate (see review by
Jungman, Kamionkowski \& Griest 1996 and references therein). In the
Minimal Super-symmetric Standard Model (MSSM), the super-partners of
the gauge bosons and the neutral Higgs bosons, respectively called
binos and higgsinos, mix into four Majorana fermionic mass
eigenstates, called neutralinos. In some regions of the super-symmetric
parameter space, the lightest neutralino can naturally achieve a relic
density that matches the observed cosmic DM abundance, thus making it
a theoretically well motivated DM candidate. The neutralino mass can
be anywhere between $\approx 50$ GeV, which is the lower bound allowed
by accelerator constraints (Eidelman et al. 2006, see {\it ibidem} for
the set of assumptions made on the underlying super-symmetric
scenario), up to 100 TeV in some extreme non-perturbative scenarios
(Profumo 2005), although masses larger than a few TeV are commonly
considered as ``unnatural'', if one wants SUSY to solve the
theoretical problems (such as the hierarchy problem) for which it was
originally invented.

An interesting alternative candidate arises in theories with Universal
Extra Dimensions (UED), and corresponds to the first Kaluza-Klein state
of the hypercharge gauge boson (Appelquist, Cheng and Dobrescu, 2002;
Servant and Tait 2000). The existence of a viable dark matter candidate
in UED theories can be seen as a consequence of the conservation of
momentum in a higher dimensional space. To generate chiral fermions at
the zero mode, the extra dimensions must be modded out by an orbifold,
leading to the conservation of the so-called KK-parity, such that all
odd-level KK particles are charged under this symmetry, thus ensuring
that the lightest (first level) KK state is stable.

A lower bound on the mass of the lightest Kaluza-Klein particle (LKP)
comes from electroweak measurements, and depending on the mass of the
Higgs, it can be as low as $\approx 300$ GeV (Gogoladze and Macesanu
2006). Although the annihilation cross section in these scenarios is
fixed by the LKP mass, it is possible to achieve the correct relic
density even for particles as heavy as several TeV, provided that one
includes the effect of co-annihilation with other KK particles
(Burnell \& Kribs 2006, Kong \& Matchev 2006, Kakizaki et al. 2006).
We note that the arguments presented here are valid more generally
for any thermal relic WIMP candidate with s-wave annihilation.

Regardless of its precise nature, it is possible to set cosmological
constraints on the WIMP mass $m_{\chi}$ under some simplifying, but
rather general assumptions on its nature. The WIMP mass is in fact
constrained to be roughly $m_{\chi}\,\gsim\,30\,{\rm GeV}\,$ and
${m_{\chi}}\,\lsim\,{10\,{\rm TeV}}$, by theoretical considerations
of the thermal freeze-out in the early Universe (e.g.  Bertone, Hooper
and Silk 2005). The very same theoretical considerations suggest that
the WIMP annihilation rate in the local Universe is far below the
expansion rate of the Universe. However, annihilations proceed in high
density DM regions, such as the centres of DM haloes. We demonstrate
in this work that this resultant annihilation luminosity could play an
important role in galaxy formation by providing a strong source of
feedback which prevents gas from cooling and forming stars in
galaxies.  WIMP pair annihilation in high DM density regions will
inevitably produce high energy neutrinos, positrons, anti-protons and
gamma-rays. There are several on-going and future experiments focused
on direct and indirect searches for the signature of neutralinos (see
reviews by Bergstrom 2002; Bertone, Hooper \& Silk 2004). In this
paper we focus on GeV mass-scale WIMPs, however for illustration
purposes we also calculate the consequence of heating by MeV DM.

\section{The self-annihilation luminosity in a simple theoretical model}
\label{section3}

Here we describe the key ingredients of the simple theoretical model
constructed to estimate the self-annihilation luminosity of DM in the
centres of galaxy and cluster scale haloes. In the concordance
$\Lambda$CDM cosmology the thermally averaged annihilation rate of
WIMPs, in absence of co-annihilations, is related to the DM relic
density by $<\!\sigma v\!>\,\sim\,{3 \times
10^{-27}}\,/\Omega_{\chi}\,h^2\,{\rm cm^3s^{-1}}$.  From this
expression one can estimate the annihilation luminosity from
neutralinos in a halo with a given density profile (assuming they
constitute all the dark matter):
\begin{eqnarray}
L_{\chi \chi} = \frac{<\!\sigma v\!>}{2\,m_{\chi}^2}\,
\int_V\,\rho_\chi^2(\vec x, t)\,d^3{x} ~, 
\label{lum}
\end{eqnarray}
where the integral is performed over the halo volume $V$, $m_{\chi}$
is the WIMP mass, and $\rho_\chi(\vec x, t)\!\approx\!\rho_\chi(r, t)
$ is the dark matter density profile, commonly assumed to be
spherically-symmetric, making it only dependent on the galactocentric
radial coordinate $r$.

\subsection{The global dark matter halo profile}

The density profile of a dark matter haloes that assemble in
cosmological N-body simulations is well fit by the following
parametrised functional form, referred to as the generalised
Navarro-Frenk-White (NFW) profile (Navarro, Frenk \& White 1996):
\begin{eqnarray}
  \rho(r)= \frac{\rho_0}{(r/r_s)^{\gamma}
  \big[ 1+(r/r_s)^{\alpha} \big]^{(\beta-\gamma)/\alpha}} ~,
\label{profile} 
\end{eqnarray}
where $\alpha$, $\beta$, and $\gamma$ govern the slope of the profile on
large and small scales respectively, with a transition in slope at the 
scale radius $r_s$, defined as the ratio of the virial radius of the
DM halo, $R_{\rm vir}$, and the virial concentration parameter, $c_{\rm vir}$:
\begin{eqnarray}
r_s = \frac{R_{\rm vir}}{c_{\rm vir}} ~.
\label{rs}
\end{eqnarray}
In this paper we only consider profiles for which the outer slope is
$-3$, i.e. where $(\beta-\gamma)/\alpha=2$.  This leads to the
following expression for the normalisation constant, $\rho_0$,
\begin{eqnarray}
\rho_0 = \frac{M_{\rm vir}}{4\,\pi\,\int_0^{R_{\rm vir}}
  \frac{r^2}{(r/r_s)^{\gamma}\,(1+r/r_s)^{3-\gamma}}{dr}} ~, 
\label{rho0}
\end{eqnarray}
where $M_{\rm vir}$ is the virial mass of the system.

The slope of the density profile in the inner regions of the halo
$\gamma$ is of interest for our work. A value of $\gamma= 1$ refers to
the NFW profile, while $\gamma= 1.5$ corresponds to the so-called Moore
profile (Moore et al. 1999). Note that eqn.~(\ref{profile}) is a fit
to the output of numerical simulations which do not include either
particle physics and hydrodynamic effects. Such effects may alter the
distribution of DM over time. The consequences of this possibility are
discussed in the next section.  In fact, it is the time evolution of
the density profile that holds the key to possibly tapping the energy
from the neutralino self-annihilation. We briefly note here that for a
cluster scale DM halo, with typical virial mass ~ 10$^{14-15}\,\msun$,
the luminosity generated from eqn.~(\ref{lum}) with central density
profile $\gamma\!=\!1\!-\!1.5$ is, at most, of order $10^{39}$
ergs$^{-1}$. Thus, even if this entire energy output coupled maximally
to the cooling gas in the inner regions it would be insufficient to
offset the cooling luminosity (of order $\sim\,10^{42-44}$
ergs$^{-1}$) in a typical cluster system (Totani 2004; 2005). Physical
processes that further steepen the inner density profile slope are
therefore required to make the annihilation viable as a feedback
mechanism. Fortunately, several astrophysical processes in the centres
of dark matter haloes are expected to affect the density profile of DM
in precisely the required fashion.

\subsection{The formation of a central density spike}

\subsubsection{Steepening due to adiabatic response to collapsing baryons}

Density enhancements can arise in the inner regions of dark matter
haloes from a number of different astrophysical processes. One such
process is the adiabatic response of the dark matter to the infall and
collapse of baryons (Blumenthal, Flores \& Primack 1986; Ryden \& Gunn
1987; Gnedin et al. 2004). Baryonic gas loses energy through radiative
processes and falls into the centre of a dark matter halo to form
stars. As a result of this redistribution of mass, the gravitational
potential of the inner regions of the halo changes. The dark matter
responds to the subsequent deepening of the potential by altering its
distribution and thereby enhancing its density. The increase in dark
matter density from adiabatic compression can be calculated using
adiabatic invariants. The cusp index $\gamma$ in the NFW profile of
eqn.~(2) steepens to a value that depends on the density profile of
the baryons.

Earlier calculations by Blumenthal et al.(1986) over-predicted the
steepening due to the assumptions of spherical symmetry and circular
orbits.  This was due, in part, to the fact that haloes in the
hierarchical structure formation scenarios grow via multiple violent
mergers and accretion along filaments, and particle orbits in the
haloes are highly eccentric. Gnedin et al. (2004) revisited this
question using high-resolution cosmological simulations that included
gas dynamics, radiative cooling, and star formation. They found that
the dissipation of gas indeed increased the density of dark matter and
steepened its radial profile in the inner parts of the halo when
compared with haloes without cooling. Comparisons with the earlier
work of Blumenthal et al.(1986) showed that the assumption of
spherical symmetry induces a systematic over-prediction of the density
enhancement in the inner 5\% of the virial radius. Gnedin et
al. (2004) correct for this by providing a simple modification of the
assumed adiabatic invariant which includes orbit-averaged particle
positions.

If the baryons have a radial density profile $\rho_b (r) \propto
r^{-\nu}$, then the spike (i) retains the same slope if $\nu = 1$;
(ii) or if $\nu > 1$, the contracted inner slope of the DM profile is
steeper than its original value. For $\nu = 1 - 2$, the final inner DM
density slope can be as steep as $\gamma = 1 - 1.7$, respectively. In
cluster sized systems, although baryons represent only a small
fraction of the overall mass, they may be crucially important on
scales comparable to the extent of the typical brightest, central
cluster galaxies.

There has been a lot of recent activity in the simulations community
to understand the likely interactions between baryons and dark matter
(Gnedin et al. 2004; Nagai \& Kravtsov 2005; Faltenbacher et al. 2005;
Gustafsson, Fairbairn \& Sommer-Larsen 2006). For instance, in a
recent simulation that included baryons, Gustafsson, Fairbairn and
Sommer-Larsen (2006) claim that the central DM cusps steepen to $\rho
\sim r^{-1.9 \pm 0.2}$, with an indication of the inner logarithmic
slope converging on galaxy mass scales. Gustaffson et al. (2006) claim
that the difference in the extent of adiabatic contraction and
subsequent response they find compared to Gnedin et al. (2004) and
other works originates in the differences in their stellar feedback
prescription. So the extent of steepening due to adiabatic response is
an unsettled issue at the present time due to the inherent
uncertainity in the our understanding and implementation of star
formation in simulations. On the observational side there have also
been many recent attempts to disentangle the dark matter and baryonic
components in clusters (Zappacosta et al. 2006; Biviano \& Salucci
2006; Mahdavi et al. 2007; Sand et al. 2008). 

Extensive convergence studies have shown that modern highest
resolution dissipationless simulations agree in their
predictions: the average logarithmic slope of the density profile at
$r = 0.01\,R_{\rm vir}$ is $\gamma = 1.3$, with a substantial scatter
of $\pm\,0.3$ from object to object (Fukushige et al. 2004; Tasitsiomi
et al. 2004; Navarro et al. 2004; Reed et al. 2004; Diemand et
al. 2004; 2005). At the same time, despite a significant decrease in
the smallest reliably resolved scale, the logarithmic slope continues
to get shallower with decreasing radius without reaching an asymptotic
value. For the purposes of this work, for our default model we will
assume a conservative value for the inner slope of the density profile
of $\gamma = 1.0$, which is the standard NFW value commonly used.

\subsubsection{Steepening due to adiabatic growth of central black
holes}

On extremely small scales $(r<1{\rm pc})$, i.e. at the very centre of
the galaxy hosted by a DM halo, the gravitational potential is
dominated not by DM (as described by eqn.~\ref{profile}) but by
baryons, mainly comprising stars and frequently a supermassive black
hole (SMBH). In fact, from the demography of nearby galaxies it
appears that nearly every galaxy hosts a SMBH, and their masses are
well correlated with properties of the stellar component in the
inner-most regions (Magorrian et al. 1998; Tremaine et al. 2002;
Ferrarese \& Merritt 2002).  In our own Galaxy, for instance, stars
exhibit a cusp inside a few parsecs described by (Genzel et al. 2003):
\begin{eqnarray}
\rho_*(r) \sim 3.2 \times 10^{5}\,{\rm
  M_{\odot}\,{pc^{-3}}}\,\Big(\frac{r}{1\,{\rm pc}}\Big)^{-1.4}~.
\end{eqnarray}
This distribution is centred around the supermassive black hole at the
Galactic centre, whose mass is estimated to lie in the range $2 - 4
\times 10^6\,\msun$ (Ghez et al. 2005; Genzel et al. 2003b).

In fact, Hooper, Finkbeiner \& Dobler (2007) claim that the excess
microwave emission from the region around the center of our Galaxy
detected by WMAP (Wilkinson Microwave Anisotropy Probe) could be
synchrotron emission from relativistic electrons and positrons
generated in dark matter annihilations.  Hooper et al. (2007) find
that the angular distribution of this ``WMAP Haze'' matches the
prediction for dark matter annihilations with a cusped density
profile, $\rho(r) \propto r^{-1.2}$ in the inner few kiloparsecs.
Comparing the intensity in different WMAP frequency bands, they find
that a wide range of possible WIMP annihilation modes are consistent
with the spectrum of the haze for a WIMP with a mass in the 100 GeV to
multi-TeV range.  Most interestingly, they find that to generate the
observed intensity of the haze, the dark matter annihilation cross
section is required to be approximately equal to the value needed for
a thermal relic, $\sigma v \sim 3 \times 10^{-26}$ cm$^3$/s.

More generally, the {\it adiabatic} growth of a massive object at
the centre of a power-law distribution of DM with index $\gamma$
induces a redistribution of the DM (also referred to as a density
`spike'), into a new power-law with a steepened index $\gamma_{\rm
spike}$ (Peebles 1972, Young 1980, Ipser \& Sikivie 1987, Quinlan et
al. 1995, Gondolo \& Silk 1999).
\begin{eqnarray}
\rho_{\rm spike}(r) = \rho(r_{\rm outer\ spike})\,\Big(\frac{r}{r_{\rm
    outer\ spike}}\Big)^{-\gamma_{\rm spike}} ~,
\label{spikeprofile}
\end{eqnarray}
where
\begin{eqnarray}
\gamma_{\rm spike} = 2 + \frac{1}{4 - \gamma}~,
\label{spikegamma}
\end{eqnarray}
and the outer spike radius is approximated by
\begin{eqnarray}
r_{\rm outer\ spike} \approx \,r_{\rm bh} \equiv
0.2\,\frac{\rm{G}\,M_{\rm bh}}{\sigma^2} ~,
\label{spikeouter}
\end{eqnarray}
where $r_{\rm bh}$ is the gravitational radius of influence of the
black hole (Merritt \& Szell 2006). The density $\rho(r_{\rm outer\
spike})$ is found by evaluating eqn.~\ref{profile} at the outer spike
radius.  For the Milky Way, $r_{\rm bh} \approx 1 {\rm pc}$.

A physical cut-off to the otherwise diverging profiles of
eqn.~(\ref{profile}) and eqn.~(\ref{spikeprofile}) at small $r$ is provided
by the self-annihilation rate itself. We can define a limiting radius,
$r_{\rm lim}$, where the density reaches a maximal value given by
\begin{eqnarray}
\rho_{\rm spike}(r_{\rm lim}) = \frac{m_{\chi}}{<\sigma v>\,t_{\rm
    spike}} ~,
\label{limdens}
\end{eqnarray}
where $t_{\rm spike}$ is the lifetime of the density spike (see
below).  This limiting radius $r_{\rm lim}$ can then be found by
equating and solving eqn.~(\ref{limdens}) and eqn.~(\ref{spikeprofile})
simultaneously. In reality, the inner cut-off radius cannot be
arbitrarily small.  Hence, we truncate the inner annihilation
luminosity at
\begin{eqnarray}
r_{\rm inner\ spike} = {\rm Max}\ [4\,R_S,\,r_{\rm lim}] ~, 
\label{rcut}
\end{eqnarray}
where $R_S$ is the Schwarzchild radius of the SMBH, $R_S =
\rm{G}\,M_{\rm bh}/c^2$.  This defines the inner spike radius, $r_{\rm
inner\ spike}$.

\subsection{Density spike evolution and enhanced annihilation
  luminosity} 

In order to understand the time evolution of a DM density spike, we
note that once the spike is established several astrophysical and
particle physics effects act to disrupt it. The three principal
effects that can damp a density spike are: (i) the self-annihilation
process itself that depletes DM (discussed above), (ii) the
interaction between the DM and the surrounding baryons near the SMBH,
and (iii) spike disruption during galaxy (or more specifically black
hole) mergers.

DM and baryons interact gravitationally with each other, and stars
typically have significantly larger kinetic energies than DM
particles. Gravitational encounters between these two populations will
tend to drive them toward equipartition of energy, causing the DM to
heat up, flattening the spike while retaining the shape of the density
profile. As shown by Bertone \& Merritt (2005a; 2005b), the resultant
spike evolution with time can be modelled as an exponential decay via
\begin{eqnarray}
\rho_{\rm spike}(r,t) = \rho_{\rm spike}(r,0)\,e^{-\tau/2} ~,
\end{eqnarray}
and
\begin{eqnarray}
r_{\rm outer\ spike} (t) = r_{\rm spike} (0) \, e^{(-\tau/2)/(\gamma_{\rm spike}-
\gamma)}~,
\end{eqnarray}
where $\tau$ is the time elapsed since the original spike formation
$t_{\rm spike}$ measured in units of the heating time $T_{\rm heat}$,
\begin{eqnarray}
T_{\rm heat} = 1.67 \times 10^{9}\,{\rm yr}\,\Big(\frac{M_{\rm bh}}{4 \times 10^
  6\, M_{\odot}}\Big)^{1/2} \,\Big(\frac{r_{\rm bh}}{2 {\rm
    pc}}\Big)^{3/2} ~.  
\label{heatingtime}
\end{eqnarray}
We show results for the time evolution of a fiducial spike in
at the centre of our Galaxy in Figure~1 and discuss the issue
further in Section~5.1.

Mergers of black holes, and in particular major mergers (i.e. SMBH
mass ratios greater than 0.3), are efficient at flattening or
destroying density spikes (Merritt et al. 2002).  We note that the
disruption is maximal during equal mass mergers, however these are
extremely rare. After a merger the density spike will be
re-established on a time-scale comparable to the relaxation time of
stars (Merritt, Harfst and Bertone 2006). If major mergers are too
frequent then spikes will not be long-lived enough to enhance the
annihilation luminosity.  However, if they never happen the spike
decay described above will result in inefficient heating.  We
encapsulate spike disruption in eqn.~(\ref{limdens}) by following the
time since the last SMBH major merger in each halo, $t_{\rm spike}$,
which approximates the spike lifetime.  The major merger rate in the
simulation and galaxy formation model is discussed further in
Section~\ref{sec:spikes}.

Using all the above computed quantities, we can finally estimate the
time dependent annihilation luminosity at any time after the most
recent SMBH major merger as:
\begin{eqnarray}
L(t;m,{\sigma v}) \, = \, {2\pi} \, \frac{<\sigma v>}{m_{\chi}^2}\,\int_{r_{\rm
    inner\ spike}}^{r_{\rm outer\ spike}}\,\rho_{\rm spike}^2(r)\,r^2\,dr
\label{heating}
\end{eqnarray}
\begin{eqnarray}
\ = \, 2\pi \, \frac{<\sigma v>}{m_{\chi}^2} \ 
\frac{\rho(r_{\rm outer\ spike})^2}{2\gamma_{\rm spike}\!-\!3} \   
r_{\rm outer\ spike}^3 \,
\Big( \frac{r_{\rm inner\ spike}}{r_{\rm outer\ spike}} \Big)^{-2\gamma_{\rm
    spike}+3}, \nonumber
\end{eqnarray}
in the limit that $r_{\rm outer\ spike}\!\gg\!r_{\rm inner\ spike}$.
In what follows, we evaluate eqn.~(14) for DM haloes in the Millennium
Run simulation.

\section{Annihilation heating in a cosmological model of galaxy
  formation}
\label{section4}

\subsection{The Millennium Run Simulation and Galaxy Formation Model}
\label{MR}

To explore the effects of the heating of cooling gas in the hot halo
from neutralino annihilation we employ a model of galaxy formation
(Croton et al. 2006) coupled to a high resolution N-body simulation,
the so-called `Millennium Run' (Springel et al. 2006). Below we give
only a brief outline of these techniques; the interested reader should
refer to Croton et al (2006) and references therein for further
details. In the following sub-sections we describe the addition of
neutralino heating to this model.

The Millennium Run N-body simulation follows the dynamical evolution
of approximately 10 billion dark matter particles in a periodic box of
side-length $500\,h^{-1}$Mpc with a mass resolution per particle of
$8.6\times 10^8\,h^{-1}{\rm M}_{\odot}$.  The adopted cosmological
parameter values are $\Omega_\Lambda=0.75$, $\Omega_{\rm m}=0.25$,
$h=0.73$, and $\sigma_8=0.9$ (Colless et al. 2001, Spergel et
al. 2003, Seljak et al. 2005).  From the simulation outputs merger
trees are constructed that describe in detail how structures grow as
the Universe evolves. These trees form the backbone onto which the
galaxy formation model is coupled. Inside each tree, virialised dark
matter haloes at each redshift are assumed to attract ambient gas from
the surrounding medium, from which galaxies form and evolve.  The
galaxy formation model effectively tracks a wide range of physics in
each halo, including reionization of the inter-galactic medium at high
redshift, including radiative cooling of hot gas and the formation of
cooling flows, star formation in the cold disk and the resulting
supernova feedback, black hole growth, metal enrichment of the
inter-galactic and intra-cluster medium, and galaxy morphology shaped
through mergers and merger induced starbursts.

More specifically, once a dark matter halo has grown in mass to
approximately $\sim\!10^{11.5} h^{-1}M_\odot$ infalling baryons no
longer fall `cold' on to the central galaxy but instead shock heat to
the virial temperature of the dark matter halo to form a quasi-static
hot atmosphere (Croton et al. 2006). As the density of hot gas
increases the cooling time near the centre of the halo becomes short
and a cooling flow of condensing gas forms. Using simple
thermodynamic and continuity arguments a cooling rate for this flow,
$\dot{m}_{\rm cool}$, can be estimated under the assumption that the
gas is approximately isothermal ($\rho_{\rm hot} \sim r^{-2}$) outside
the very central regions at the virial temperature of the halo
($T_{\rm vir} \sim V_{\rm vir}^2$, where $V_{\rm vir}$ is the virial
velocity of the dark halo):
\begin{equation}
\dot{m}_{\rm cool} = 0.5\, 
\Big(\, \frac{r_{\rm cool}}{R_{\rm vir}}\, \Big)\, 
\Big(\, \frac{m_{\rm hot}}{t_{\rm cool}}\, \Big) ~. 
\label{cooling}
\end{equation}
Here $R_{\rm vir}$ is the virial radius of the halo, $m_{\rm
hot}$ is the mass of gas in the hot phase, and $t_{\rm cool} \approx
0.1\,t_{\rm Hubble}=R_{\rm vir}/V_{\rm vir}$ is the cooling time of
the gas at the cooling radius, which is defined by:
\begin{equation}
r_{\rm cool} =  \Big[ \frac{\Lambda(T,Z)}{6\pi \mu m_{\rm p} kT}
\frac{m_{\rm hot}}{V_{\rm vir}} \Big]^{1/2}~,
\label{rcool}
\end{equation}
The cooling radius traditionally marks the radius out to which gas has
had time to cool quasi-statically given the age of the system.  In the
above, $\mu m_{\rm p}$ is the mean particle mass, $k$ is the Boltzmann
constant, and $\Lambda(T,Z)$ is the cooling function, dependent on
both the temperature $T$ and metallicity $Z$ of the hot gas. Despite
their simplicity, eqns.~(15) and (16) provide a good approximation to
the rate at which gas is deposited at the centre in the similarity
solution for a cooling flow (Bertschinger 1989; Croton et al. 2006).

Gas cooling in simulations appears to be much more efficient than
observed in nature.  For low mass haloes heating and expulsion of
baryonic material by supernovae can prevent over-cooling, but on group
and cluster scales, while the observed baryon fraction has the
universal value (Balogh et al. 2001), state of the art hydrodynamical
simulations find a significantly larger fraction of cooled gas
(Borgani et al. 2003; Kravtsov, Nagai \& Vikhlinin 2005). In fact in
clusters the problem is particularly severe; observationally the X-ray
luminosities in many clusters imply cooling rates in the inner regions
of $\sim$ 100 - 1000 ${\rm \msun\,yr^{-1}}$ (e.g. Fabian \& Nulsen
1977; Edge 2001; Fabian 2004), whereas little or no associated star
formation is detected. A decrease in X-ray gas temperature is often
seen in the centres of clusters but X-ray spectroscopy detects very
little cool gas below $10^7$ K at temperatures of $\sim 1 - 2\,{\rm
keV}$ (Peterson et al. 2001; 2003).

Therefore it has been known for some time (see Fabian 1994 and
references therein), the measured cooling times at the centre of
observed clusters of galaxies are significantly shorter than the
Hubble time.  This implies that gas condensation should be occurring
from the conventional picture described above, however none is
observed, nor are the secondary effects of cooling such as copious
star formation in the central brightest cluster galaxy. Such galaxies
are mostly found to be `red and dead'. This behaviour may indicate the
presence of a heat source that is supplying energy to the cooling gas
to keep the central regions from condensing.  This is the so-called
classical `cooling flow problem' (Fabian 1994).

To arrest the expected runaway cooling in massive systems a heat
source must be present. We test the idea of whether the heating from
dark matter annihilation at the centres of massive systems can
plausibly be a global solution to the cooling flow problem described
above. Another key aspect of the cooling flow problem is the fact that
the cooling time of gas is short over a fairly large range of radii,
particularly in groups and clusters, implying that the heating rate
of any proposed mechanism ought to balance the cooling at all radii.
Below we describe our implementation of this DM annihilation
powered heating into the galaxy formation model.

\subsection{Thermal coupling of the neutralino annihilation energy}

The energy produced from the self-annihilation of neutralinos in the
vicinity of the spike $r_{\rm spike}$ we argue will couple thermally
with the baryons in galactic nuclei, thereby heating the gas. We
investigate potential physical processes and the relevant time-scales
that will likely enable the transfer of this energy to the baryons. We
compute relevant time-scales for the centre of the Milky Way treating
that as our fiducial case from which we scale to other masses. Similar
heurisitic estimates have been made by Totani (2004; 2005) for galaxy
clusters.

{\bf The dominant astrophysical process via which the
self-annihilation energy interacts with baryons is unknown at the
present time.} Here for purposes of estimating the efficiency of
potential astrophysical processes we use observational estimates of
the density, pressure and temperature from the inner regions of our
Galaxy at roughly 200 pc from the Galactic Center (Bradford et al.
2005; Morris \& Serabyn 1996). The physical conditions present in this
region are consistent with thermal, non-thermal and magnetic pressures
that are several orders of magnitude higher than those present in the
large-scale galactic disk. For the measured densities of $n \sim
10^{4}\,{\rm cm^{-3}}$, the pressure is $P_{\rm thermal} \sim
10^{-10}\,{\rm erg\,cm^{-3}}$, however turbulent pressures greatly
exceed this value, approaching $P_{\rm turb} \sim 10^{-8}\,{\rm
erg\,cm^{-3}}$. For the inferred magnetic field strength of $0.1 - 1$
mG, the magnetic pressure is also large, of the order of $P_{\rm mag}
\sim 10^{-8} - 10^{-10}\,{\rm erg\,cm^{-3}}$. To estimate the
time-scales for energy loss and therefore heating of the gas in the
inner regions, we adopt the following scaling values for the density,
pressure and magnetic field strength from the Milky Way: $P \sim
10^{-8}\,{\rm erg\,cm^{-3}}$, $n \sim 10^{4}\,{\rm cm^{-3}}$, and $B
\sim 0.4$ mG. {\bf Note that the self-annihilation heating occurs for
the $T \sim 10^4$ K cooling gas from the hot halo and not to the $T
\sim 200$ K molecular gas at 200 pc.}

Using the above values we can address the question of how the
annihilation luminosity is likely converted efficiently into thermal
energy of the gas in the nucleus. We focus on a fiducial neutralino
model where about $1/4$th of the annihilation energy $E_{\chi\chi}$
goes into continuum gamma-rays [Channel A], $1/6$th goes into
electrons and positrons [Channel B], $1/15$th goes to protons and
anti-protons [Channel C], and the rest is imparted to neutrinos that
are not relevant for heating. The spectral energy distribution of the
products is such that $E^2\,\frac{dN}{dE}$ peaks roughly at
$0.05\,m_{\chi}\,c^2$, $0.05\,m_{\chi}\,c^2$ and $0.1\,m_{\chi}\,c^2$
respectively. The average energy of secondary particles produced in
the annihilation of a neutralino of mass 50 GeV is of the order of
$1\,{\rm GeV}$.  These annihilation products are representative of a
wide class of super-symmetric scenarios, as determined with the
DarkSUSY code (Gondolo et al. 2005), that are consistent with
cosmological and accelerator constraints.

Examining Channel B in detail, we assume for simplicity that all the
electrons and positrons have roughly the same energy $E_0 \sim 1$ GeV.
Since they are produced in an extremely high density environment, they
will expand relativistically and form a bubble. As discussed in
Shu (1991), the primary energy loss processes are
examined below. The energy loss time-scale for these particles by
Coulomb collisions is given by:
\begin{eqnarray}
\tau_{\rm CC} \sim 5.1 \times 10^3\,\frac{n}{10^4\,cm^{-3}}\,{\rm E_0\,yr},
\end{eqnarray}
which is so short that all the electrons and positrons from the
neutralino decay will very efficiently transfer energy to the gas
cooling out of the hot halo. This is the principal physical process
that will likely heat the gas and provide the feedback discussed in
this work.  On the other hand, the energy loss time from inverse
Compton scattering is long,
\begin{eqnarray}
\tau_{\rm IC} \sim 1.2 \times 10^9\,{\rm E_0}^{-1}\,{\rm yr},
\end{eqnarray}
and therefore inefficient. The time-scale for energy losses via
Brehmstrahlung is also short,
\begin{eqnarray}
\tau_{\rm Brehm} \sim 5.7 \times 10^3\,\Big(\frac{n}{10^4\,{\rm
    cm^{-3}}}\Big)^{-1}\,{\rm yr}.
\end{eqnarray}
In comparison using the same estimate for $n$, the energy loss time-scale from synchrotron radiation is
given by,
\begin{eqnarray}
\tau_{\rm Sync} \sim 1.5 \times 10^5\,{\rm
  E_0}^{-1}\,{\rm yr} 
\end{eqnarray}
We conclude from the above estimates that the conversion of the
self-annihilation energy into thermal energy can occur efficiently for
the gas in the inner regions of galactic nuclei as illustrated
specifically for the Milky Way case. However, we do note here that our
estimate of the magnetic field strength is on the high side. This energy input suppresses the
gas cooling and quenches star formation.  Due to our lack of knowledge
of neutralino properties, it is not possible to calculate the precise
time-scale and physical process that thermally couples the by-products
of the $\gamma$-ray annihilation to baryons. For the purposes of this
work, it is assumed that this coupling is efficient and all the
available energy is transferred effectively to heating the gas in the
inner regions of DM haloes. The estimate of time-scales for the Milky
Way halo suggests that gas in the inner region on the scale of
$\sim\,200\,{\rm pc}$ is likely directly involved in the heating
process.

We note here that an alternate dark matter driven gas heating
mechanism exploiting inelastic scattering of X-dark matter particles
with relic abundances comparable to neutralinos has been proposed by
Finkbeiner \& Weiner (2007). They propose a WIMP candidate with an
excited state that maybe collisionally excited and de-excites by $e^+
- e^-$ pair emission. The kinetic energy of these pairs they argue
could heat intra-cluster gas and the gas in galaxies to compensate for
cooling similar to what we explore here. The primary motivation for
this model was to provide a possible interpretation for the 511 keV
line observed by the INTEGRAL satellite in the inner Milky Way
consistent with the observed WMAP haze and current constraints on the
gamma-ray background.

\subsection{Constructing a viable feedback model from neutralino
  self-annihilation} 

We model the self-annihilation luminosity of neutralinos as a steady
input of energy that prevents gas cooling in preferentially massive
galaxies/haloes. The assumptions and properties of simulated haloes
that are needed to fully specify the DM density profile, BH mass, and
the inner and outer spike radii are all taken as input by the galaxy
formation model to calculate the annihilation luminosity for any
system at any given time its evolution.

\subsubsection{Halo density profiles} 

The DM density profile for every halo in the Millennium Run at every
time-step is approximated using the universal analytic function
described by eqn.~(\ref{profile}). Note that, while in principle each
profile can be directly measured from the distribution of bound dark
matter particles, this is computationally prohibitive given the number
of haloes in the Millennium run (up to 25 million at any given
time-step), and the analytic formalism is accurate enough for our
purposes here.

To explore the dependence of annihilation heating on the inner slope
of the halo density profile, $\gamma$, we take the inner slope as a
free parameter in determining eqn.~(\ref{profile}).  In the next
section we will vary $\gamma$ and examine its effect on the evolution
of the galaxy population.  The remaining parameters in the density
profile are either assumed fixed or taken directly from the
simulation.

For the default model, we assume $\alpha=1$ and choose $\beta$ such
that $(\beta-\gamma)/\alpha=2$.  This ensures that the outer slope of
the halo density profile remains fixed at $-3$, which is known to be
an accurate description of the results of numerical simulations
(Navarro, Frenk \& White 1996).  We take the measured virial mass
$M_{\rm vir}$, and radius $R_{\rm vir}$, required in eqn.~(\ref{rs})
and (\ref{rho0}), directly from the simulation. To estimate the halo
concentration parameter $c_{\rm vir}$, we use the measured $V_{\rm
vir}$ and $V_{\rm max}$ and solve eqn.~(5) in Navarro et al. 1997 (see
Croton, Gao \& White 2007).  This fully describes the density profile
for each dark matter halo.

\subsubsection{The inner and outer spike radii}

Once a value for the slope of the inner dark matter density profile,
$\gamma$, has been assumed, the steepened spike index, $\gamma_{\rm
spike}$ (eqn.~\ref{spikegamma}), can be calculated.  This then fixes
the spike density profile defined by eqn.~(\ref{spikeprofile}), and
therefore also the inner limiting density determined by the
self-annihilation rate itself, eqn.~(\ref{limdens}).  The radius of
this limiting density, and the inner spike radius, are calculated from
eqn.~(\ref{spikeprofile}) and (\ref{limdens}).  Note that we limit the
inner spike radius to always be equal to or greater than four times
the Schwarzchild radius (eqn.~\ref{rcut}).  However, this limit is
rarely reached in practice.

The outer spike radius is simpler to calculate for each halo in the
galaxy formation model, since the model explicitly follows the growth
of SMBHs in each galaxy (Croton et al. 2006). This, along with the use
of the virial velocity of the halo $V_{\rm vir}$ as a proxy for the
inner velocity dispersion $\sigma$, allows the outer spike radius to
be calculated using eqn.~(\ref{spikeouter}).

\subsubsection{The efficiency of annihilation heating}

We consider a maximal heating model, assuming that all the energy
available from the annihilation luminosity given by
eqn.~(\ref{heating}) couples with the cooling gas in the hot halo.
With this assumption, the cooling rate described by
eqn.~(\ref{cooling}) is modified in the presence of DM annihilations:
\begin{eqnarray}
{\dot m_{\rm cool}}^{'} =  {\dot m_{\rm cool}} - \frac{\xi\,L_{\chi
    \chi}}{1/2\,{V_{\rm vir}}^2} 
\label{finalheating}
\end{eqnarray} 
Thus, if the heating rate from the annihilation flux is comparable to
the energy released from gas cooling out of the hot X-ray halo, the
cooling flow can be suppressed and this will starve the central galaxy
from lack of new star forming material. Under such circumstances
galaxy growth will stall, altering the relationship between dark halo
mass and galaxy luminosity in a way more compatible with observations.

\subsubsection{Uncertainties in the galaxy formation model}

As the results in the following sections are considered, it is
important to keep in mind that the galaxy formation model we use to
obtain them is imperfect.  Its construction is largely based on
observational phenomenology that, while well described in the mean, is
often ill understood in detail.

The largest uncertainty in our galaxy formation model relevant to our
results is the cooling prescription described in Section~\ref{MR}. The
physics of cooling losses from hot plasma in a dynamically evolving
multiphase medium is complicated.  Our prescription provides cooling
rates that are a reasonable average approximation to that obtained
from hydrodynamically simulated gas infall (see Yoshida et al. 2002
for a comparison).  However, cooling in both the hydrodynamic
simulations and our galaxy formation model is first calculated in the
absence of any heating (eqn.~\ref{cooling}). The model heating rate is
then subtracted (eqn.~\ref{finalheating}).  But in reality, cooling and
heating will occur simultaneously.  Energy injection from heating can
potentially modify the properties of the gas (notably temperature and
density) used to calculate later cooling.  Hence, under these
circumstances subsequent cooling estimates will be different from
those where the past heating history is neglected as in our model.

Due to these caveats, in absolute terms our heating rates must be
treated with caution; they are simply a relative measure that defines
an upper limit on the energy required to stop gas from
cooling. However, relative to the cooling rate they are expected to be
accurate for the model.  More detailed models can be constructed to
take into account the past heating history when calculating the
current cooling rate, but these estimates would be complicated and
highly unconstrained.  For example, the temperature and density gas
profiles could change as a function of the energy injected, depth of
potential, and redshift, for every galaxy in an evolving Universe.
These detailed and self-consistent responses of the baryons need to be
taken into account. However, such improvements add considerable
complexity that is beyond the scope of the present work. Below, we
present the results of implementing a simple model of annihilation
heating.

\section{Results}
\label{section5}

\subsection{Spike evolution and stability}
\label{sec:spikes}

As discussed in Sections~3.1 and 3.3, the luminosity produced from
dark matter annihilation in an unsteepened NFW-type halo falls short by
several orders-of-magnitude when compared to the cooling losses from
the X-ray emitting hot gas. For annihilations to be a plausible
solution to the cooling flow problem one requires an enhancement of
the DM density in the inner most regions of the galaxy, usually
assumed to be produced by the presence of a super-massive black hole
or in response to the adiabatic compression of baryons or perhaps
both these processes.

\begin{figure}
\includegraphics[width=\columnwidth]{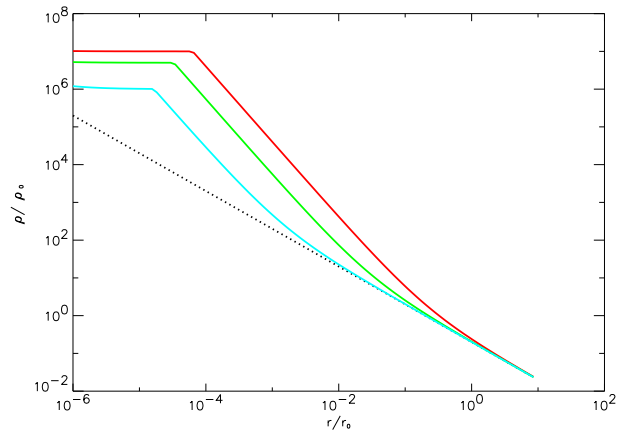}
\caption{To illustrate the time evolution of the DM spike, we plot the
variation of the density in the centre with time for a DM halo that
hosts a Milky Way type galaxy. The x-axis is normalised to $r_h$ the
radius of gravitational influence of the black hole, while the y-axis
is normalised to $\rho_0$ the density of the DM halo at $r_h$. The 3
different line styles denote the values of $\tau$ (i.e. time in units
of the heating time $T_{\rm heat}$). The solid lines are, from top to
bottom, for $\tau=1$ (red), $\tau=5$ (green) and $\tau=10$ (blue).
The "unmodified" NFW profile is shown as a dotted line. The central
"plateau" for the curves is due to annihilations, and the maximum
density is calculated for a toy model, to simply show the qualitative
behaviour of the profile at small radii. The maximum density - the
height of the plateau, varies with time.}
\label{fig:spike_evol}
\end{figure}

When present, density spikes are expected to evolve with time due to
the annihilation itself.  To illustrate this, in
Fig.~\ref{fig:spike_evol} we plot the time evolution of a DM density
spike in the centre of a fiducial halo, modelled on the Milky Way. The
x-axis is normalised to the radius of gravitational influence of the
black hole, while the y-axis is normalised the density of the DM halo
at this radius. The three different line styles denote the values of
$\tau$ (i.e. time in units of the heating time $T_{\rm heat}$,
eqn.~\ref{heatingtime}). The red line is for $\tau=1$, green for
$\tau=5$, and blue for $\tau=10$. The unmodified NFW profile
(i.e. $\gamma=-1$) is shown as a dotted curve. The central plateau in
the inner regions is due to annihilations, and the maximum density
drops with time. The evolution of the density spike is clearly a
result of many complex interactions within the halo.

Dark matter density spikes are fragile and transient -- a major merger
can easily destroy a spike on a short time-scale. However, it is
unclear whether the disruption is triggered by the merger of the 
dark matter haloes, central black holes or galaxies hosted in the
haloes. The merger rates for these three populations of objects are
not necessarily the same, and yet a typical approximation adopted is
to the use Extended Press-Schechter (EPS) theory to obtain the merger
rate. However EPS only describes the properties of DM haloes. In
reality it is probably the mergers of black holes that ought be most
significant for the evolution of a density spike.

\begin{figure}
\includegraphics[width=\columnwidth]{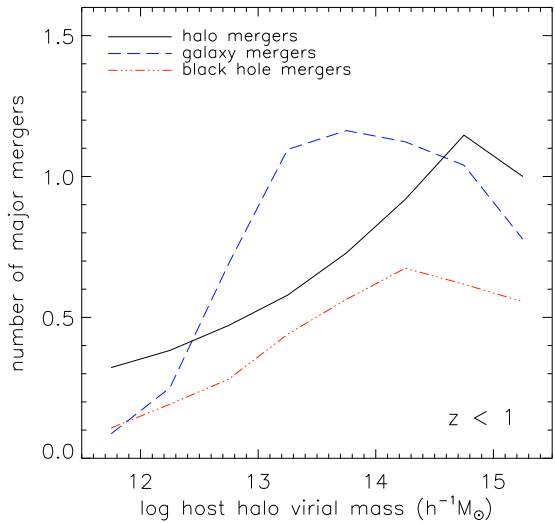}
\caption{The number of major mergers for galaxies, DM haloes, and
black holes vs host halo mass. This plot illustrates that the density
spikes can be reassembled as the time between major mergers is long
compared to the time taken to refill the stellar loss cone in galactic
nuclei.}
\label{fig:merger_rate}
\end{figure}

Using our cosmological model of galaxy formation we can check the
frequency of mergers for haloes, galaxies, and the black holes that
reside within them. In Fig.~\ref{fig:merger_rate} we show for each
object the mean number of major mergers at $z\!<\!1$. We focus on this
redshift range because it is primarily only at late times wherein
heating is needed to prevent the cooling of gas in haloes.  Major
mergers of SMBHs are less common than major mergers of galaxies or
dark matter haloes, with less than one occurring per system on average
since $z\!=\!1$ at all host halo mass scales. This implies a typical
spike survival time $>\!4\,{\rm Gyr}$.  Fig.~\ref{fig:merger_rate}
also shows that the merger rates for all populations tends to increase
with increasing mass although this flattens somewhat for the most
massive haloes. We conclude from Fig.~\ref{fig:merger_rate} that
spikes can survive on Gyr timescales despite the violent nature of the
hierarchical growth and assembly in a $\Lambda$CDM Universe.  Note,
that even after a major merger spikes may later reform, typically on
the stellar relaxation time-scale of the inner region of the galaxy.

In this work, we further assume that the mass of central black holes
grows significantly at each merger (as the solution for adiabatic
growth in valid only in the limit $M_{\rm bh, initial}\,<<\,M_{\rm bh,
final}$ (Di Matteo, Springel \& Hernquist 2005).

\subsection{Annihilation heating of the hot halo}
\label{sec:heating}

\begin{figure*}
\includegraphics[width=\textwidth]{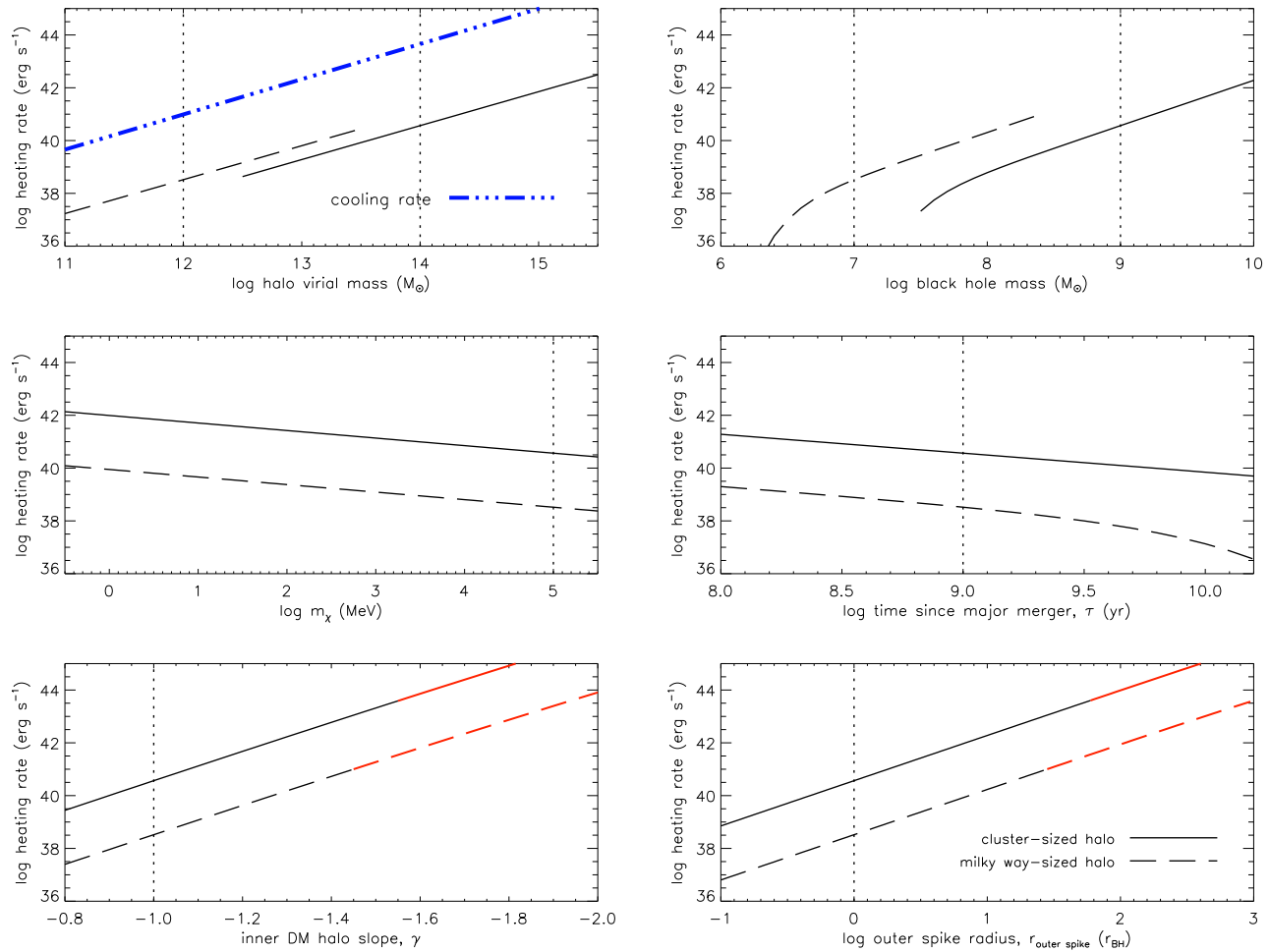}
\caption{The parameters controlling the feedback from
self-annihilation: the mass of the dark matter halo (top left), black
hole mass (top right), DM particle mass (middle left), time since
major merger (middle right), inner DM halo slope (bottom
left), and outer spike radius (bottom right). Two models are shown,
one for a Milky Way sized halo and one for a cluster sized halo.  The
default values for each are shown by the dotted horizontal lines in
each panel.  The common parameters are $m_{\chi} = 100\,{\rm GeV}$,
$\tau = 1\,{\rm Gyr}$, $\gamma = -1.0$ (NFW), and $r_{\rm
outer\,spike} = r_{\rm BH}$ (see text).  Each panel demonstrates the
effect on the heating of varying one of these parameters while keeping
the remaining fixed. The annihilation luminosity is most sensitive to
the inner halo density profile and outer spike radius.  In the top
left panel we additionally illustrate the approximate cooling rate for
a given halo mass that the heating needs to overcome (thick
dotted-dashed line, assuming a hot gas fraction of 10\%.).}
\label{fig:heating}
\end{figure*}

We now investigate the enhanced annihilation luminosity due to the
presence of density spikes described in Section~\ref{section4}. The
annihilation heating model depends on a well defined set of
parameters.  We explore the dependences in detail. These dependencies
are illustrated in the various panels of Fig.~\ref{fig:heating}: the
dark matter halo virial mass ${M_{\rm vir}}$, the black hole mass
${M_{\rm bh}}$, the mass of the dark matter particle ${m_{\chi}}$, the
average time between major mergers $\tau$, the inner DM halo density
profile slope $\gamma$, and the outer radius ${r_{\rm outer\,spike}}$
adopted for the DM density spike. We consider two halo mass ranges for
each, galaxy scales masses ($M_{\rm vir}=10^{12}M_\odot$, dashed
lines) and cluster masses ($M_{\rm vir}=10^{14}M_\odot$, solid lines).
Our default model assumes common values for each of $m_{\chi} =
100\,{\rm GeV}$, $\tau = 1\,{\rm Gyr}$, $\gamma = -1.0$ (NFW), and
$r_{\rm outer\,spike} = r_{\rm BH}$ (the sphere of influence of the
black hole). In each panel we vary one parameter keeping the rest
fixed to clearly show the plausible range of heating rates that can be
expected from the model (the default values of each parameter are
marked by vertical dotted lines). In addition, the top left panel
shows the expected cooling rate assuming a hot gas fraction of
0.1. The choice of 10\% for the host gas fraction is motivated by the
X-ray observations in galaxy groups and clusters (LaRoque et al. 2006;
Vikhlinin et al. 2006; Gonzalez, Zaritsky \& Zabludoff 2007). This is
approximately the energy the DM heating needs to replenish.

It is clear that even in the presence of density spikes it is usually
difficult to produce enough flux to reheat all the cooling gas, and
this is true for haloes of all masses ranging from galaxy scales to
cluster scales.  Enhancement of between 1-2 orders-of-magnitude does
occur when mergers are more frequent (this works to offset the
exponential decay of the spike amplitude shown in
Fig.~\ref{fig:spike_evol}) or for lower mass DM particle
candidates. However the heating luminosity is still significantly
lower when compared with the cooling rate shown in the top left panel.

The bottom two panels of Fig.~\ref{fig:heating} show how sufficient
heating can be produced, either through significant steepening of the
inner DM halo density profile (on which the spike sits), or a
significantly larger outer spike radius or both. For galaxy-scale DM
haloes annihilation heating can balance cooling whenever the inner
density profile is steepened to values of $\gamma\,<\,-1.45$, and for
cluster-scale haloes when $\gamma\,<-1.55$. Similarly large values for
the outer spike radius, $\sim\,25\,r_{\rm bh}$ on galaxy scales and
$\sim\,65\,r_{\rm bh}$ on cluster scales, are required for
annihilation heating to compensate for the gas cooling.  Although
observationally the inner DM halo profile remains unconstrained on
extremely small scales, it may be difficult to produce such steep
slopes or large spike radii on the scales required for all galaxies
and clusters, in order to effectively influence global properties.

Another challenge for our model is the requirement that heating and
cooling balance needed at all radii as the cooling time for the gas is
short for a range of radii specially on group and cluster scales. As
developed here self-annihilation dumps all the energy at very small
radii and here we assume due to lack of a more complete understanding
at the present that all that energy is also thermalized efficiently at
small radii. While this could lead to the generation of large entropy
inversions in the X-ray gas that are typically not observed (Voit \&
Donahue 2005); we argue that there are other efficient energy
transport processes like thermal conduction available to deposit
energy at larger radii (Ruszkowski \& Begelman 2002; Ruszkowski,
Bruggen \& Begelman 2004). It is useful to point out here that even
for the alternative AGN feedback that have been proposed the
astrophysical process that couple radiation to baryons in the inner
regions of galaxies, groups and clusters is poorly understood at
present (McCarthy et al. 2008). This is clearly a rich avenue for
future work for all currently proposed feedback models.

\subsection{The consequences of annihilation heating in a cosmological
  context} 

\begin{figure*}
\includegraphics[scale=0.6]{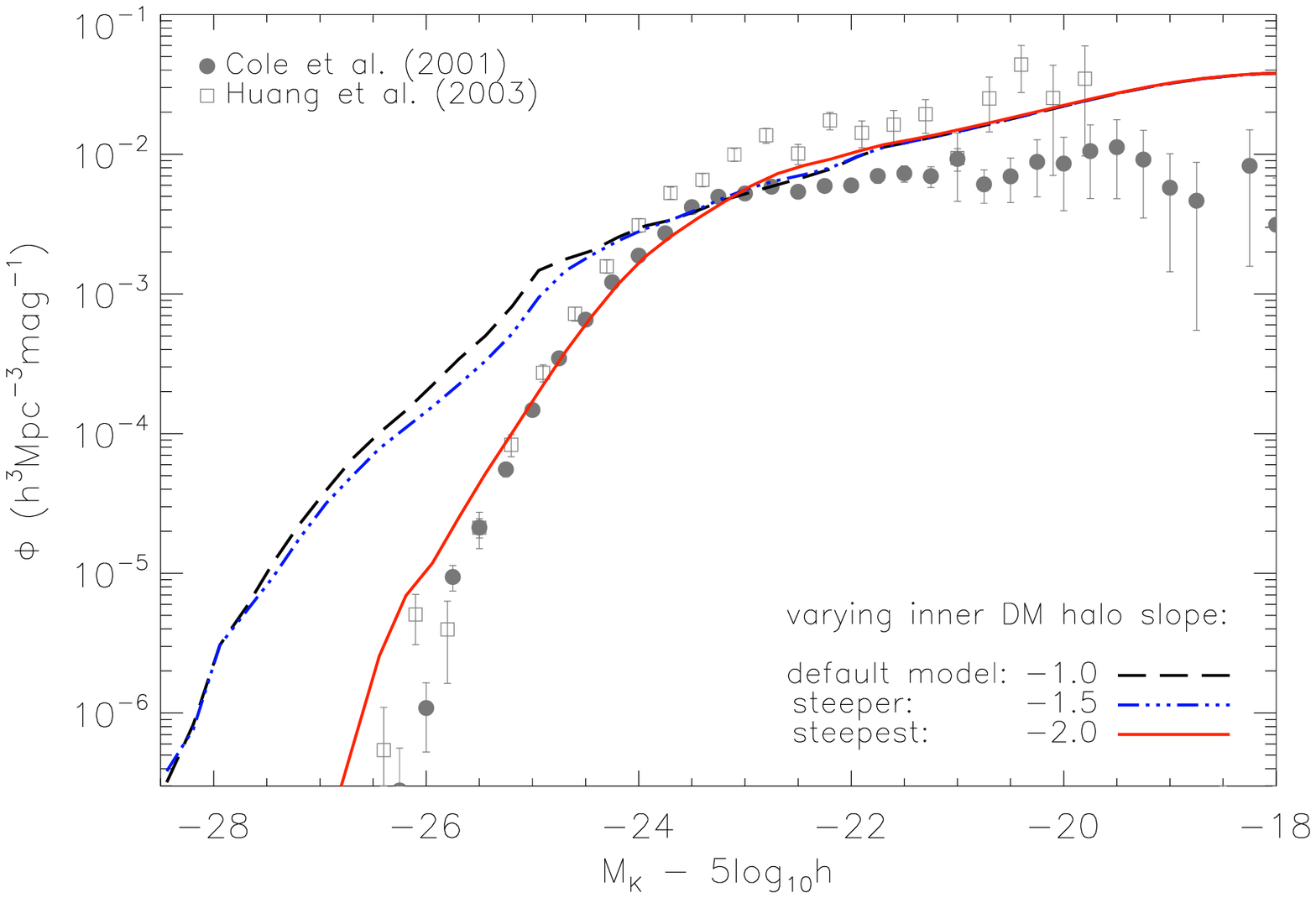}
\caption{The local K-band galaxy luminosity function, observed
(points with error bars), and predictions (lines) for various
parameter choices of our annihilation heating model.  Here we show the
default model of Fig.~\ref{heating} with $\gamma = -1.0$, and with
steepening DM density slopes of $\gamma = -1.5$ and $\gamma = -2.0$,
as indicated in the legend.  Only spikes in haloes with the 
steepest inner profiles are able to produce enough heating to obtain a
reasonable fit to the observed galaxy luminosity function.}
\label{fig:LFs1}
\end{figure*}

\begin{figure*}
\includegraphics[scale=0.6]{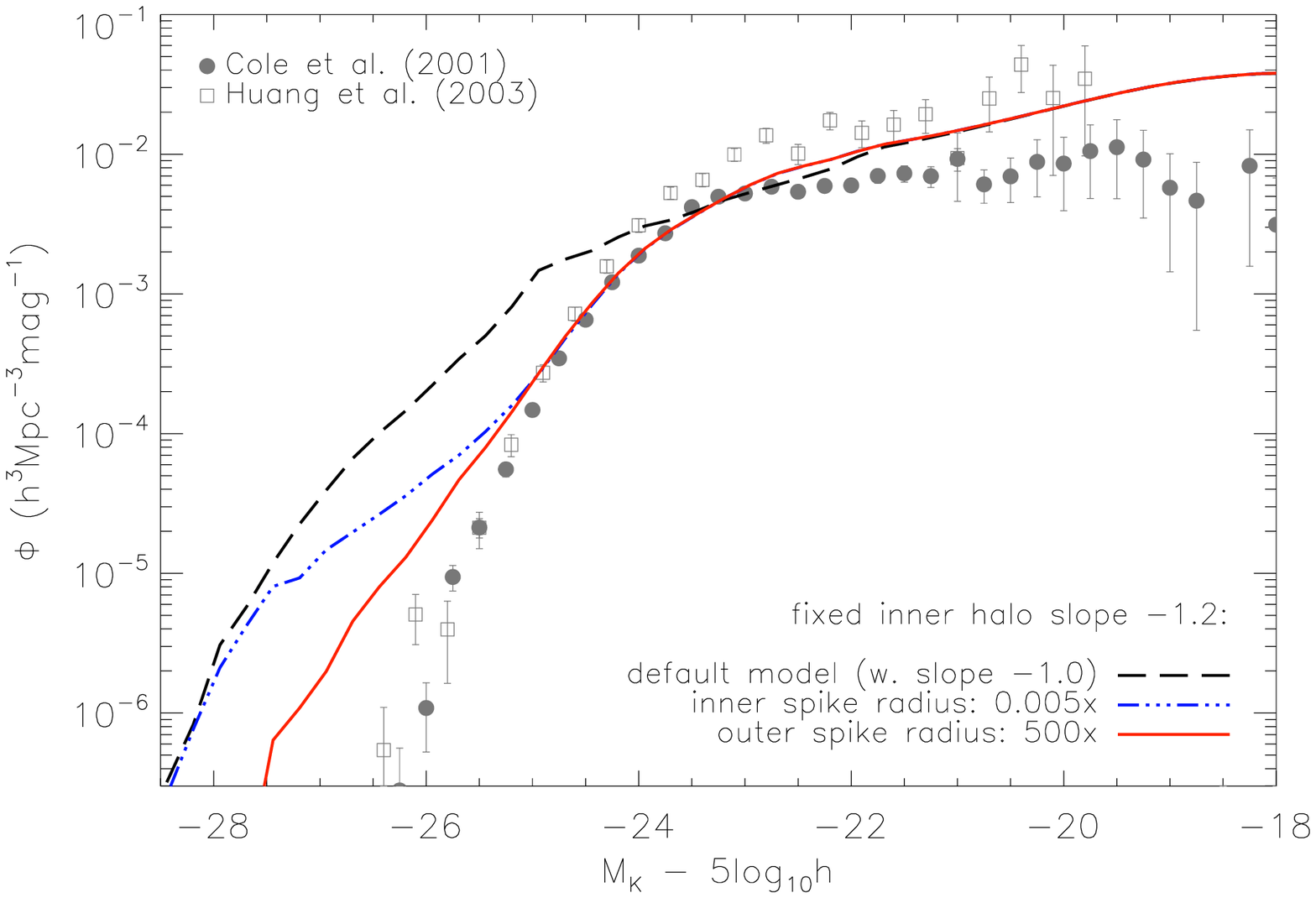}
\caption{The local K-band galaxy luminosity function, observed (points
with error bars), and predictions (lines) for various parameter
choices of our annihilation heating model.  The default model with
inner slope $\gamma = -1.0$ is replotted from Figure~\ref{fig:LFs1}.
For comparison, haloes with an inner slope of $\gamma = -1.2$ and more
extreme choices for the inner and outer spike radii are shown, as
indicated by the legend.}
\label{fig:LFs2}
\end{figure*}

\begin{figure*}
\includegraphics[scale=0.6]{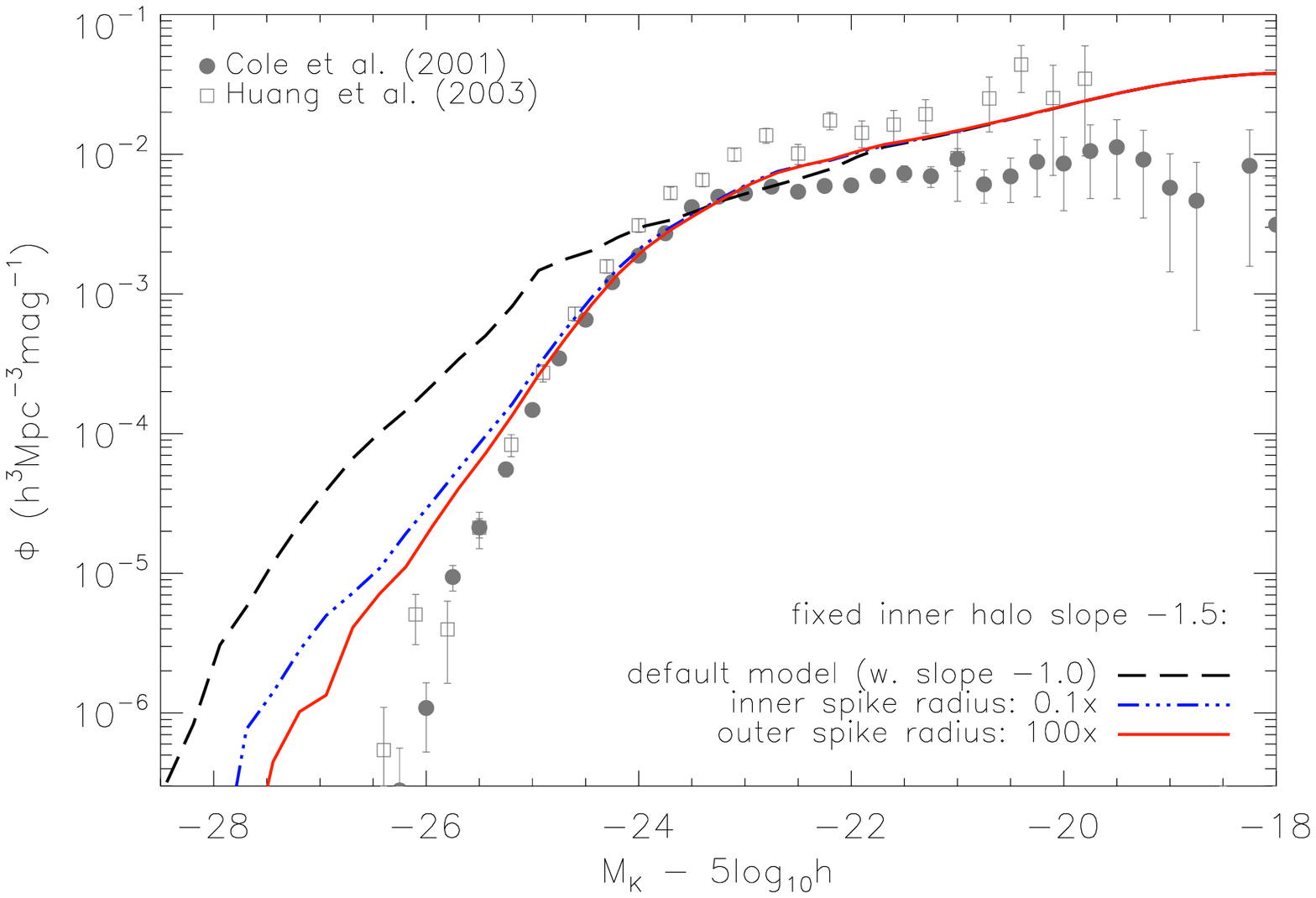}
\caption{The local K-band galaxy luminosity function, observed (points
with error bars), and predictions (lines) for various parameter
choices of our annihilation heating model.  The default model with
inner slope $\gamma = -1.0$ is replotted from Figure~\ref{fig:LFs1}.
For comparison, haloes with an inner slope of $\gamma = -1.5$ and more
extreme choices for the inner and outer spike radii are shown, as
indicated by the legend.}
\label{fig:LFs3}
\end{figure*}

The suppression of cooling flow gas in massive haloes can have a
dramatic effect on evolution of the galaxies that reside in them.
This is due to the fact that across cosmic time, a growth of a galaxy
is dictated by the availability and supply of star forming material.
At late times this mostly comes in the form of gas condensing out of
the hot halo.  Hence, any mechanism that suppresses gas cooling
ultimately also prevents the galaxy from further star formation. Our
goal is to investigate under what circumstances the evolving density
spikes explored in Section~\ref{sec:spikes} and the resultant heating
model described in Section~\ref{sec:heating} can actually shut down
star formation in massive galaxies when the full hierarchical
evolution of galaxies is taken into account.

In a set of three figures, Fig.~\ref{fig:LFs1}--\ref{fig:LFs3}, we
show the resultant local luminosity function predicted by our galaxy
formation model with annihilation heating included.  We plot the $z=0$
K-band galaxy luminosity function, for the models (lines) and
observations for comparison (symbols with error bars). The long-dashed
line in each figure shows the conservative default model used in
Fig.~\ref{fig:heating}. The remaining lines in each figure illustrate
the consequences of different parameter choices for the annihilation
heating prescription, with each figure focused on a specific set of
parameters that tune the annihilation heating model.

As can been seen from all the luminosity function figures, the default
model significantly overpredicts the abundance of the brightest
galaxies.  This is a consequence of inefficient annihilation flux
heating for the default model, as seen in
Fig.~\ref{fig:heating}. Essentially overcooling occurs in the centers
of group and cluster systems, leading to excess star formation and
overly bright and massive galaxies. Note that this is not necessarily
a failure of our underlying galaxy formation model - the cooling flow
problem has a long history (see review by Fabian 2004 for details) and
failure of the default heating model is simply another manifestation
of it (in the absence of strong enough heating).

The two additional lines in Fig.~\ref{fig:LFs1}, dotted-dashed and
solid, show the galaxy luminosity function when the inner DM halo
slope is steepened to $\gamma=-1.5$ and $\gamma=-2.0$, respectively
(the default model has $-1.0$, the standard NFW profile). An inner
slope of $-1.5$ appears to be insufficient to produce enough heating,
even with the presence of a central density spike. Only inner halo
slopes of $\sim -2.0$ or steeper are able to do this. In the context
of current models such steep slopes do not arise naturally in dark
matter haloes. A combination of steepening mechanisms needs to operate
in a coordinated fashion to achieve these slopes.

Fig.~\ref{fig:LFs2} considers the combination of a steeper inner halo
slope, here $\gamma = -1.2$, and more extreme values for the inner and
outer spike radii explored previously in Fig.~\ref{fig:heating}.  Our
choices for the spike radii are made to obtain the correct turnover in
the galaxy luminosity function.  The required values are, for the
inner spike radius $200$ times smaller than the default value
(dashed-dotted line) and for the outer spike radius $500$ times larger
than the default value (solid line).  Both parameter sets still over
predict the abundance of very bright galaxies.  This exercise is
repeated in Fig.~\ref{fig:LFs3} with a more extreme inner slope of
$\gamma = -1.5$.  The inner and outer spike radii choices that provide
the best fit are now $10$ times smaller than the default inner spike
radius (dashed-dotted line) and $100$ times larger than the default
outer spike radius (solid line).  The brightest galaxies remain
overabundant, but less so than for the previous values.

In Figs.~ 7 \& 8 we plot the heating rate and cumulative heating rate respectively 
generated by dark matter self-annihilation as a function of radius for a galaxy scale
halo and a cluster scale halo for various parameter choices. Also marked as arrows 
for reference in these plots are 3 relevant physical scales:  4 $R_S$ (inner-most arrow);
the outer spike radius (middle arrow) and the virial radius of the DM halo (outer-most arrow).
The plots emphasize that the equations predict that the heating flux comes from well 
inside the cooling radius. Due to the simplicity of our assumed cooling model (i.e. that
the gas is isothermal) it is not possible to accurately determine the cooling rate as a
function of radius inside individual halos for a direct radial comparison of the heating
cooling balance. In the context of our simple model as shown we can however calculate the global 
heating and cooling rates and compare. These plots also show that the heating from 
self-annihilation akin to other modes of feedback needs to propagate out to larger radii 
to be effective. The microphysics of these transport processes is poorly understood at this point.

We conclude that for modest parameter choices we are unable to produce
galaxies that match observed ones. Of course, further combinations of
these parameters are possible, however, it is unlikely that
self-annihilation is the sole feedback process in galaxy formation. It
is plausible that this process operates in addition to supernovae and
AGN feedback.

\begin{figure*}
\includegraphics[scale=0.8]{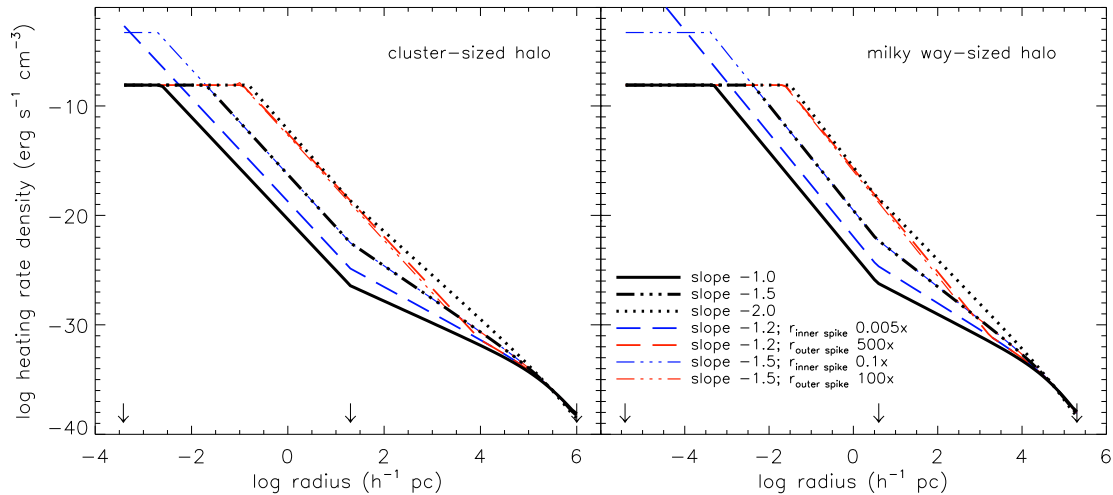}
\caption{The dark matter heating rate density profile for the various
heating models explored in Figures~\ref{fig:LFs1} to \ref{fig:LFs3},
for both cluster-sized and Milky Way-sized haloes (left and right
panels). The three arrows near the bottom of each panel indicate the
boundaries of $4$ times the Schwarzchild radius, the outer spike
radius, and the virial radius of the DM halo, from left to right in
each respectively (see Section~\ref{section3}).  The spike is clearly
noticable as the inner steepening part of the profile, while the
flattening seen on small scales marks the saturation of the heating
due to DM self annihilation.}
\label{fig:radial}
\end{figure*}

\section{Observational signatures of DM annihilations}

If indeed neutralino self-annihilations contribute to the energetics
of feedback in galaxies and clusters as proposed here, we can expect
a range of observational signatures. 

\subsection{Distribution of density profile slopes}

If the model of annihilation heating developed here operates, we
expect the existence of a range of inner dark matter density profile
slopes in the centres of galaxies and galaxy clusters. The time
evolution explored in this model suggests that the inner density
slopes on the scale of tens of parsecs in galaxies are likely to be
diverse.

Observationally, this is an extremely challenging length scale to
probe and detect this diversity. Studies of the velocity dispersion
profiles of the stellar component in the vicinity of the black holes
combined with strong lensing offer a potentially viable probe.

Mapping of rotation curves has also provided constraints on the
density profile of the dark matter in the central regions of galaxies,
(see de Blok, McGaugh \& Rubin and references therein) however on much
larger scales, of the order of kpc, whereas the DM annihilation
scenario leaves an imprint on much smaller scales. 

It does appear on that on kpc scales (larger scales than relevant for
dark matter self-annihilations) there is compelling evidence for
bimodality in the distribution of light (baryons). The NUKER group has
studied this effect extensively using Hubble Space Telescope data
(Gebhardt et al. 1996; Faber et al. 1997; Lauer et al. 2002). In a
recent paper, combining several HST investigations on the central
structure of early-type galaxies they find that the distribution of
the logarithmic slopes of the central brightness profiles is bimodal
(Lauer et al. 2007). They claim that at the HST resolution limit, most
galaxies are either power-law systems, which have steep cusps in
surface brightness, or core systems, which have shallow cusps interior
to a steeper envelope in the brightness distribution. There is a
atrong correlation between the luminosity $L$ and inner profile slope,
and it has been suggested that this correlation is likely due to core
formation by binary BHs during mergers (Ferrarese et al. 2006).
Whether and how this observed bimodality in the surface brightness
profiles of the baryonic component reflects the dark matter density
profile on the smallest scales is unclear at the present.

The physical scale on which dark matter annihilation manifests itself
in the case of clusters is predicted to be of the order of $\sim$ kpc
(as shown in the bottom right hand panel of Figure~3). In this
context, we predict that similarly in clusters there ought to be a
diversity of density profile slopes on kpc scales. In clusters that
have more complex dynamical histories, the dark matter spike is likely
to have been disrupted progressively due to frequent mergers and these
density spikes are also expected to have depleted from the
annihilation process itself as a result. These growing clusters are
systems in which the spike reassembly is most unlikely to occur
rapidly. In the context of the self-annihilation feedback picture,
these clusters are likely to have density profiles with a central
plateau (akin to the evolution shown in Figure~1). Since clusters are
the most recently assembled structures in the Universe, we predict a
range of inner density slopes in the central few kpc, some shallower
than the predictions of dissipationless simulations and some steeper,
depending on their dynamical history. Dynamical history coupled with
the modification produced due to the presence of the baryonic
component (stars or black holes) is intricately coupled to the process
of DM annihilation as we have shown, and the interplay of these
process might dictate the slope of the dark matter density profile in
the inner-most regions.

Observationally, the issue is once again challenging. Strong lensing
studies of the inner regions of clusters with radial and tangential
arcs point to the possible existence of shallower density slopes and
perhaps cores on scales of $\sim\,5 - 10$ kpc. This is of the order of
the scales on which we expect to see signatures of the annihilation
process. Since strong lensing constrains the total mass as a function
of radius, disentangling the effect of the baryons to infer the
density profile of the dark matter alone on these scales is difficult.
The combination of gravitational lensing and dynamical data is
uniquely capable of achieving this. Sand et al. (2002; 2004) attempted
this for a sample of strong lensing clusters. In more recent work,
Sand et al. (2008) study 2 clusters Abell 383 and MS2137-23 combining
strong lensing constraints with stellar velocity dispersion data for
the brightest central cluster galaxy. They find that a shallower inner
slope is preferred compared to predictions from simulations
($\gamma\,\sim\,{-0.6}$) for Abell 383 for a coarse lensing model. For
MS2137-23, no self-consistent model that incorporates strong lensing
and the stellar velocity dispersion data was found to be a good
fit. We also note that constraints derived from recent Chandra
observations also suggest shallow inner slopes ($\gamma \sim -1$) on
scales of $\sim 5 - 10\,{\rm kpc}$ (Vikhlinin et al. 2006; Zappacosta
et al. 2006; Voigt \& Fabian 2006). It is however a real challenge to
extract robust constraints on the density profile slope at these small
radii from observations. Meanwhile, simulations that incorporate
baryons are likely to improve in resolution in the near future and
might offer a powerful test of our predictions.

\begin{figure*}
\includegraphics[scale=0.8]{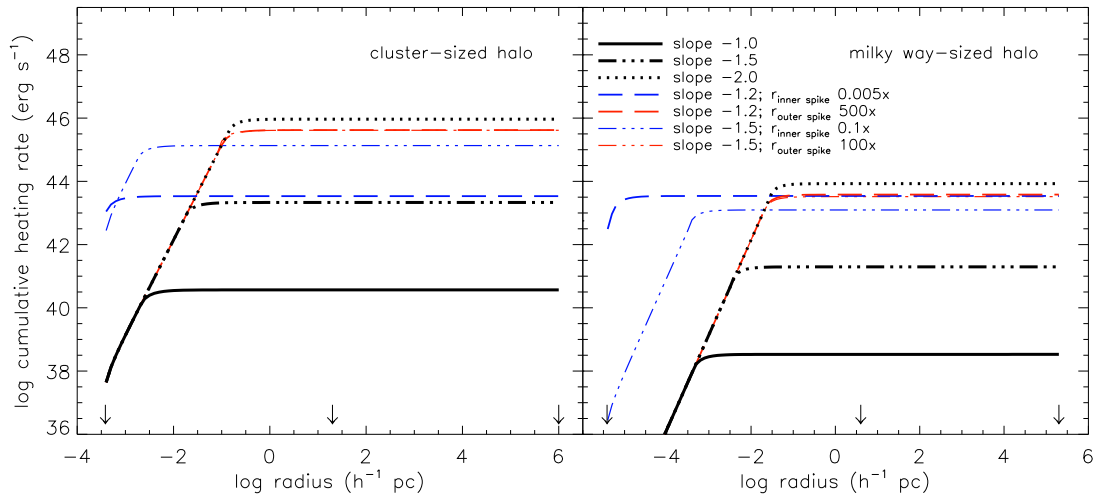}
\caption{The cumulative heating rate as a function of increasing
radius for the various spike density profiles shown in
Figure~\ref{fig:radial} and used in Figures~\ref{fig:LFs1} to
\ref{fig:LFs3}. In both panels, for cluster-sized haloes on the left
and Milky Way-sized on the right, the three lower arrows indicate the
boundaries of $4$ times the Schwarzchild radius, the outer spike
radius, and the virial radius of the DM halo, from left to right in
each respectively (see Section~\ref{section3}). For all models the
majority of the annihilation flux originates from well inside the
central most parts of the halo.}
\label{fig:radialCumulative}
\end{figure*}

\section{Comparison with the AGN heating paradigm}

In this section we discuss the details of the AGN heating paradigm in
order to contrast with our dark matter annihilation model. We describe
the mechanism and argue that there is a need for additional sources of
feedback that dark matter self-annihilations may well provide.

The proposed physical mechanisms for AGN driven feedback are expected
to occur in two modes: the `quasar mode', where mergers trigger
fuelling to central black holes activating an episodic, bright quasar
phase accompanied by large-scale outflows (Di Matteo et al. 2005;
Hopkins et al. reference), and the `radio mode', which refers to
steady feedback from low-level AGN activity (Croton et al. 2006; Bower
et al. 2006). The need for these processes arise from the
over-cooling problem that occurs on a range of mass scales, from
galaxy groups to clusters (Somerville \& Primack 1999; Cole et
al. 2001; Kauffamnn et al. 1999; Benson et al. 2003).

AGN feedback in the quasar mode likely occurs during the epochs of
efficient cold gas feeding to the central black holes via a thin
accretion disk, at high accretion rates ranging between $0.1 -
1\,L_{\rm Edd}$.  This phase is short-lived and the thermal coupling
of AGN energy is fairly weak (at $<$ 5\% level).  In this mode the
AGN-driven wind removes residual gas at the end of the merger, leading
to suppression of subsequent star formation and self-regulated BH
growth that reproduces the observed $M_{\rm bh}- \sigma$ relation
(Springel et al. 2005). However, for a typical massive galaxy in the
local Universe a quasar event was long ago in its history.  Such
galaxies instead host BHs that are accreting at much lower rates; in
fact most spend much of their lifetime in these radiatively inefficient
states. Radio activity is associated with these low accretion rate
states and radio jets are seen in many massive galaxies. The coupling
of jet energy with host gas can be very efficient and models of
effervescent heating with a combination of sound waves, weak shocks
and bubbles can heat a large fraction of the gas in clusters
(Ruszkowski et al. 2004; Churazov et al. 2001; Bruggen et al. 2005)
and produce features that match X-ray observations (Fabian et
al. 2005).

A detailed exploration of the theoretical consequences of the steady
radio-mode feedback from low luminosity AGN has been presented in
Croton et al. (2006).  This was done using the same semi-analytic
model used in this paper, but with DM annihilation heating replaced
with radio-mode heating.  The authors of Croton et al. showed that for
a set of energetically and observationally plausible parameters such a
model could simultaneously explain: (i) the low observed mass drop-out
rate in cooling flows; (ii) the exponential cut-off at the bright end
of the galaxy luminosity function; and (iii) the fact that the most
massive galaxies tend to be bulge-dominated systems in clusters and
found to contain systematically older stars than lower mass galaxies.

In a recent paper Best et al. (2007) study a sample of radio-loud AGN
in nearby groups and clusters from the Sloan Digital Sky Survey
(SDSS).  Using observational estimates of the mechanical output of
radio jets, they estimate the time-averaged heating rate associated
with recurrent radio source activity for all group and cluster
galaxies. They find that within the cooling radius the radio-mode
heating associated with galaxy groups and low mass clusters is
sufficient to offset the cooling flow from the extended hot halo.  In
the most massive brightest cluster galaxy systems, however, radio mode
heating alone is not enough. They conclude that other processes acting
in massive clusters must also be contributing to the suppression of
cooling flow gas.

Importantly for this work, although AGN appear to be making an
observable contribution to the evolution of gas dynamics in dark
matter haloes, alone they only comprise part of the full physical
picture. The presence of SMBHs at the centres of massive systems that
drive AGN winds also provide enhancement in the annihilation rate of
DM that, under the right circumstances, can produce sufficient heating
flux to arrest cooling gas. Composite DM annihilation and AGN heating
models will be a natural extension of our work and there is need for
all feedback mechanisms to be better understood. One of the current
challenges for all models including the self-annihilation proposed here
is developing a better understanding of the micro-physics of how to
thermally couple to the gas at all radii where the heating and cooling
are required to balance for feedback to be effective.

\section{Summary}
\label{section6}

Using the Millennium Run N-body simulation coupled with a
sophisticated model of galaxy formation that includes the heating of
cooling flow gas through neutralino annihilation, we have shown that:

\begin{itemize} 
\item Density spikes that support the annihilation flux at the levels
required to offset cooling flows are stable enough over long enough
time-scales to maintain a reasonably constant heating source
(Fig.~\ref{fig:merger_rate}).

\item Models that appear to be extreme at the present time (given our
current understanding of DM density profiles) are required to produce
enough heating flux to offset the predicted cooling rates. To obtain
the required heating rates we either need to steepen the inner DM
density slope to values $\gamma\,>\,1.5$ or increase the outer spike
radius. For galaxy sized haloes the outer spike radius is required to
be of the order of $\sim\,25\,r_{\rm bh}$, and for cluster haloes
$\sim\,60\,r_{\rm bh}$ (Fig.~\ref{fig:heating} and \ref{fig:LFs1}).

\item The efficiency of heating in the DM annihilation model scales
with halo mass (and circular velocity), therefore this mechanism does
provide preferential suppression of star formation in more massive
haloes as required to explain current observations of the luminosity
function (Fig.~\ref{fig:heating}).
\end{itemize}

In this treatment, we have assumed that the mass of central black
holes grows significantly at each merger (as the solution for
adiabatic growth in valid only in the limit $M_{\rm bh,
initial}\,<<\,M_{\rm bh, final}$. We note here that an alternative
source of annihilations may be provided by mini-spikes around
inter-mediate mass black holes as suggested by Bertone, Zentner \&
Silk (2005).

It is clear that feedback and energy injected into the inter-stellar
medium of galaxies and the intra-cluster medium is a complex process,
and that a combination of astrophysical processes, including the one
explored here, are likely at play. One of the key uncertainties in the
model explored in this paper arises from the fact that we lack an
understanding of the physics through which the annihilation flux is
expected to couple with the cooling hot halo gas. While we have
discussed some possibilities, like coulomb collisions, brehmstrahlung
and synchrotron radiation.  Colafrancesco et al. (2007) have explored
these in more detail for the case of the heating of gas in the Coma
cluster due to DM annihilations. Experimental confirmation of
super-symmetery from the LHC at CERN might throw new light on the
viability and likely couplings for the neutralino. Additionally while
following the cumulative heating history is very challenging to do it
is needed to really understand the detailed energetics of the
gas. There is incontrovertible evidence for the presence of copious
amounts of DM on all scales in the Universe, so DM annihilation is an
inescapable phenomenon. However, how much energy is released in the
process and how efficiently it couples to the baryonic component are
unclear.

\section*{Acknowledgements}

PN acknowledges the use of the Yale ITS High Performance Computing
facility. DC acknowledges support from NSF grant AST00-71048. The
Millenium Simulation was carried out by the Virgo Supercomputing
Consortium at the Computing Center of the Max-Plank Society in
Garching. The simulation data is publicly available at
``http://www.mpa-garching.mpg.de/Millenium''. The authors thank David
Spergel, Joel Primack, David Merritt, Douglas Finkbeiner and Sergio
Colafrancesco for useful comments on the manuscript.

\end{document}